# When Human-Computer Interaction Meets Community Citizen Science

Yen-Chia Hsu and Illah Nourbakhsh
The Robotics Institute, Carnegie Mellon University, Pittsburgh, PA, U.S.A.
{yenchiah,illah}@andrew.cmu.edu

Human-computer interaction (HCI) studies the design and use of interfaces and interactive systems. HCI has been adopted successfully in modern commercial products. Recently, its use for promoting social good and pursuing sustainability, known as *sustainable HCI*, has begun to receive wide attention [4]. Conventionally, scientists and decision-makers apply top-down approaches to lead research activities that engage lay people in facilitating sustainability, such as saving energy. We introduce an alternative framework, *Community Citizen Science* (CCS), to closely connect research and social issues by empowering communities to produce scientific knowledge, represent their needs, address their concerns, and advocate for impact. CCS advances the current science-oriented concept to a deeper level that aims to sustain community engagement when researchers are no longer involved after the intervention of interactive systems.

## Defining Community Citizen Science

Citizen science refers to the framework that empowers amateurs and professionals to form partnerships and produce scientific knowledge jointly. A major science-oriented strand aims to facilitate scientific research and address large-scale problems that are infeasible for experts to tackle alone [11]. In this strand, professionals lead projects, define the goals, and encourage the public to participate in scientific research. One example is *Galaxy Zoo*[1], which uses the knowledge collected from volunteers to classify a large number of galaxies online. Another example, *eBird*[2], is an online crowdsourcing platform that engages birdwatchers to contribute bird data collaboratively. Projects in this strand, such as Galaxy Zoo and eBird, are typically designed to answer scientific research questions and increase public understanding of science.

Community Citizen Science is a particular case of citizen science that embraces participatory democracy, community co-design, and power rebalance. These characteristics correspond to three different issues: core value, participation model, and governance structure. Depending on who defines the research question, citizen science projects can have different scientific, educational, social, environmental, and political values [11]. These projects can make use of diverse participation models between scientists and citizens, ranging from crowdsourcing to co-creation [5]. Citizen science can also apply different governance structures to connect stakeholders, ranging from top-down to bottom-up [2].

---

[1] Galaxy Zoo: https://galaxyzoo.org
[2] eBird: https://ebird.org

Participatory Democracy

CCS embraces participatory democracy to influence policy-making and address local concerns that community members wish to advocate for themselves. This community-oriented strand seeks to generate scientific evidence from community data to support exploration, understanding, and dissemination of local concerns [8]. In CCS, community members frame the main research question, and scientists engage in local issues that are raised by communities. For example, the *Bucket Brigades* project, pioneered by *Global Community Monitor*[3], provides low-cost devices that enable affected residents to collect air samples, send these samples to laboratories for analysis, measure the impact of local industrial pollution, and launch advocacy efforts. When connected to CCS, sustainable HCI extends scientific research into community empowerment, exploring how to use technology to strengthen the link between science and civil society.

Community Co-design

CCS embraces community co-design to develop interactive systems with advocacy groups, who are deeply grounded in local cultures and can bring diverse expertise to inform the design and use of computational tools [3]. In this way, CCS intends to rebalance technological privilege and develop a shared understanding of how technology is embodied in context, which brings community members and scientists into parity. Previous work has shown that a strong partnership among scientists and citizens has great potential to prompt decision making and produce policy changes [11]. For instance, The *Community-Driven Environmental Project* co-designed its technology platform, *NatureNet,* with naturalists and community members to successfully support local watershed management, such as engaging residents in installing rain barrels [9]. When connected to CCS, sustainable HCI further explores how researchers take on the role of supporters that facilitate utilizing and disseminating technology, instead of supervisors that oversee and control the entire community engagement procedure.

Power Rebalance

CCS aims to rebalance power by using a bottom-up and multi-party structure, where local communities play significant roles in initiating grassroots movements, providing organizational networking, and disseminating critical findings to influence policy-making. CCS is especially beneficial when lay perspectives contradict professional ones, and thus activism is needed to inform decision-makers about the perceptions of community concerns. In this way, CCS promotes ongoing political discourse around local concerns to improve the conditions of society. For instance, in a study of childhood leukemia cases that were clustered near contaminated water wells in Woburn, residents recruited epidemiologists to show the relationship between the risk of childhood leukemia and the hazardous chemicals in their drinking water [1]. When connected to CCS, sustainable HCI is extended to exploring how technology can empower citizens to produce scientific evidence and rebalance power relationships among stakeholders.

---

[3] Global Community Monitor: https://gcmonitor.org

## Architecting Interactive Systems as Technology Infrastructure

Designing interactive systems to support Community Citizen Science suffers from the dilemma of *Wicked Problems* [10]. These problems have no precise definition, cannot be fully observed at the beginning, are unique and depend on context, have no opportunities for trial and error, and have no optimal or provably correct solutions. While researchers intend to enable citizens to generate scientific evidence and express their concerns with interactive systems, they are unable to accurately predict if citizens will contribute sufficient data to draw meaningful insights. It is also difficult to determine the critical amount of human effort, time, and the geographical scale required for extracting reliable knowledge. Moreover, there are various methods of collecting, presenting, analyzing, and using the data. It is not feasible to explore and evaluate all possible methods without deploying the system in a real-world context. Furthermore, each context requires customized community outreach strategies due to different power relationships among stakeholders. These challenges, combined with other social conditions, make it tough to integrate modern sensing devices and computational tools to support sustainability and community empowerment.

To tackle Wicked Problems, we propose an approach inspired by the design process used in architecture and urban planning. When approaching Wicked Problems on community or city scales, architects and urban planners design physical infrastructure based on prior empirical experiences to sustain human activities without their continuous supervision. This mindset treats interactive systems as an ongoing technical infrastructure that sustains communities over time, even when the researchers are no longer present. For instance, in *Civic Technoscience*, citizens use open-source tools to conduct scientific research, raise awareness of local issues, and influence policy-making [12]. The *Balloon Mapping* project, pioneered by *Public Lab[4]*, provides low-cost technology for communities to create high-resolution landscape imagery with various applications, such as documenting protest events and evaluating the effectiveness of bioswales in absorbing pollutants. In this case, the Balloon Mapping tool became the technology infrastructure that supports community activism without extensive ongoing expert assistance following deployment.

## Evaluating the Impact of Interactive Systems

When evaluating interactive systems after deployment, we believe it is more beneficial to ask "Is the system influential?" instead of "Is the system useful?" Merely focusing on usability testing metrics, such as the time of completing tasks, may restrict the perspective of system design. Metrics of impacts, such as changes of attitude and behavior, can be useful proxies for evaluating the effectiveness of technology interventions.

---

[4] Public Lab: https://publiclab.org

However, due to the dilemma of Wicked Problems, it is extremely challenging to statistically verify whether the interactive system truly empowers communities and causes attitude or behavior changes. Unlike observational studies, Community Citizen Science applies technology to simultaneously produce scientific knowledge and influence community members. If researchers frame the question of identifying the causal relationships as an observational study, it is difficult to track and control confounding factors that may influence their behaviors and attitudes, such as the effect of news and social media.

Alternatively, evaluating the intervention of technology with a randomized experiment can suffer from scientific and ethical concerns. It is not practical to randomly sample a control group with sufficient size from affected residents, since the information can spread among communities. Even if there is a way to prevent the control group from accessing the information about the deployed system, it is not appropriately ethical and contradicts the value of democratizing scientific knowledge. One may further consider an ethical way to compare the changes in the targeting community with another independent one that shares similar concerns but does not have access to the tool at the beginning. Nevertheless, CCS is by nature not replicable since each community has distinct characteristics and power relationships. The results obtained by conducting a randomized experiment on two independent communities can be misleading.

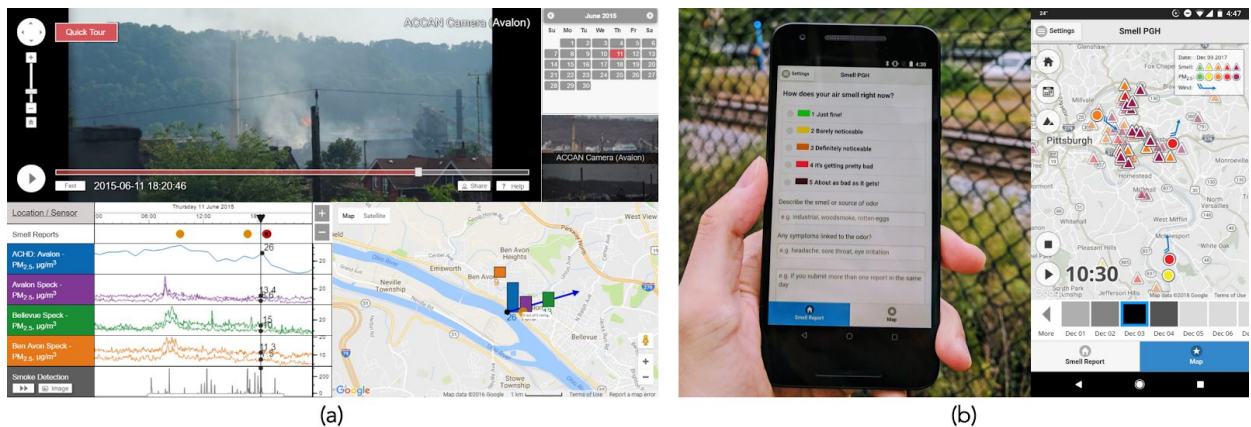

Figure 1: The left image (a) shows the user interface of our prior work, a community-empowered air quality monitoring system [6]. The right image (b) shows the user interface of our other prior work, Smell Pittsburgh [7].

Although it is difficult to statistically and rigorously validate the impact of systems on communities, understanding "How can the system be influential?" and "Does the community think that the system is influential?" can be beneficial. Through qualitative and quantitative analysis, the evaluation of impact can provide valuable insights to inform system design for the HCI community. For instance, our work, a community-empowered air quality monitoring system, enabled affected residents to present evidence of local air pollution for social activism, including smell reports, sensor data, and videos from monitoring cameras [6]. It studied how community members used animated smoke images and found that both manual and automatic approaches for generating images are essential during the engagement lifecycle. The data provided by the system, combined with personal stories from affected residents, urged regulators to respond to the air pollution problem during a public community meeting[5]. Despite the small sample size in the analysis of self-efficacy and sense of community, the survey study found that the capability of using data-driven evidence from multiple perspectives is an important reason that the communities felt more confident after interacting with the system.

Our other work, *Smell Pittsburgh*[6], is a mobile application that enables residents to report pollution odors and track where these odors are frequently concentrated [7]. From our previous work, we found that smell experiences were valuable in representing how local air pollution affected the living quality of communities. In this work, we translate this lesson from a hyperlocal to a city-wide scale and provide insight into how smartphones, sensors, and statistical methods can be used to support CCS. The data analysis showed that events of poor smell were related to a joint effect of wind information and hydrogen sulfide. A survey study found that motivations for community members to use Smell Pittsburgh came mainly from internal factors, including the desire to contribute data-driven evidence, concern about the welfare of others, and the ability to validate personal experiences using the visualization. This result was reinforced by the quantitative analysis of system usage, which identified a moderate association between contributing data and interacting with the visualization. The reports collected via Smell Pittsburgh were printed and presented by the community at the Board of Health meeting with the local health department, which urged the regulator to enact rigorous rules for coke plants[7].

## Next Steps

Community Citizen Science aims to empower everyday citizens and scientists to represent their voices, reveal local concerns, and shape more equitable power relationships. Lessons that are learned from hyperlocal scale projects may be translated to large scale ones. However, when conducting CCS research, it is essential to acknowledge that replicating successful experiences and "parachuting" technology intervention without thoughtful consideration can cause irreversible harm to communities. Developing interactive systems to support CCS requires training of multidisciplinary skills, including system development, interaction design, data analytics, community engagement, and research methods. To connect science tightly to local concerns, we call for establishing CCS as a formal research field and integrating both computational and design thinking skills into curricula, which involves understanding local concerns through community fieldwork, forming the research question, co-designing technology infrastructure with communities, and evaluating the social impact after system deployment. In this way, we go beyond the mindset of "citizens as scientists" to "scientists as citizens." We envision that CCS can drive sustainable HCI toward citizen empowerment at a time when community concerns, sustainability issues, and technological ethics are at the forefront of global social discourse.

---

[5] Regulators reviewing Shenango Coke Works' compliance with 2012 consent decree: https://www.post-gazette.com/news/environment/2015/11/19/Regulators-reviewing-Shenango-Coke-Works-compliance-with-2012-consent-decree/stories/201511190230

[6] Smell Pittsburgh: https://smellpgh.org

[7] Allegheny County Health Department defends air quality efforts, plans stricter coke plant rules: https://archive.triblive.com/local/pittsburgh-allegheny/allegheny-county-health-department-defends-air-quality-efforts-plans-stricter-coke-plant-rules/